# Against the Norm: Modeling Daily Stock Returns with the Laplace Distribution

David W. Toth[†], Bruce Jones[‡]

[†]*University of Notre Dame, Notre Dame, IN 46556*
[‡]*TAAG Indiana Wealth Management, South Bend, IN 46615*

Modeling stock returns is not a new task for mathematicians, investors, and portfolio managers, but it remains a difficult objective due to the ebb and flow of stock markets. One common solution is to approximate the distribution of stock returns with a normal distribution. However, normal distributions place infinitesimal probabilities on extreme outliers, but these outliers are of particular importance in the practice of investing. In this paper, we investigate the normality of the distribution of daily returns of major stock market indices. We find that the normal distribution is not a good model for stock returns, even over several years' worth of data. Moreover, we propose using the Laplace distribution as a model for daily stock returns.

## Introduction

It is of great interest to many investors, portfolio managers, and others to model the returns of their investments, especially stocks. However, like any financial modeling problem, fitting a model to stock returns (i.e. daily returns) is difficult, as the patterns of returns can drastically change as a result of many internal and external influences on the stock price or the market more generally. We are faced, then, with the task of deriving a flexible, yet accurate model that factors in both the long-term, more stable trends and the short-term, hard-to-predict trends.

An easy solution is to ignore the outliers and focus on the more probable events. Using a normal, or Gaussian model for daily stock returns is one such solution. Many introductory-level courses teach students to use the normal approximation for daily stock returns when modeling over a long period of time, but even some of the most influential names in mathematical finance agree that daily returns fall outside the realm of the normal distribution.[1] While it is true in many cases that the distributions of daily stock returns tend to be symmetrical, the distributions also have fat-tails, meaning there is a greater likelihood of observing extreme cases (Figure 1).

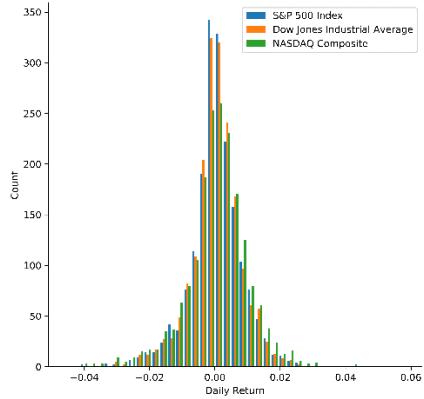

*Figure 1 - Histogram of Daily Returns of Major Stock Indices.* Data was taken from Yahoo Finance and extends back to January 2012. Note that there are multiple observations near the tails of the distributions.

In this paper, we examine the distributions of the daily returns of three major stock market indices – the S&P 500 Index, the Dow Jones Industrial Average Index, and the NASDAQ Composite Index. Using well-known measures of normality, we find that the distributions of daily returns are not Normal. Moreover, we suggest that the Laplace, or double-exponential distribution is a more suitable model for daily returns, as it is symmetric with fat-tails.

# Methods

## Calculation of Daily Returns and Laplace Distribution

Though a logarithmic return approach is commonplace in financial modeling, we calculate and use the simple daily return in this paper. Let $R_t$ be the simple return of the stock at time $t$, $P_t$ be the price or value of a stock/index at time $t$, and $P_{t-1}$ be the price or value of a stock/index at time $t-1$. Then,

$$R_t = \frac{P_t - P_{t-1}}{P_{t-1}}$$

We also clarify the definition and parameters of the Laplace distribution that we use to model the daily stock returns. A random variable $X$ has a Laplace distribution if its probability density function is:

$$f(x) = \frac{1}{2\lambda} e^{-\frac{|x-\mu|}{\lambda}}$$

where $\mu$ is the center of the distribution, and $\lambda$ is the scale of the distribution. Following literature, we estimate the center with the sample median and the scale with the mean absolute deviation from the median:

$$\hat{\lambda} = \frac{1}{n}\sum_{i=1}^{n} |x_i - \hat{\mu}|$$

### Skew, Kurtosis, and Shapiro-Wilk

To numerically evaluate the normality of daily stock returns, we turned to the third and fourth moments of a probability distribution – skew and kurtosis – as well as to dependable statistical tests such as the Shapiro-Wilk test for normality. If a random variable $X$ has a probability density function $f(x)$, then the skew and kurtosis are computed via the integrals[2]:

(Skew) $\quad \mu'_3 = E[X^3] = \int_{-\infty}^{\infty} x^3 f(x) dx$

(Kurtosis) $\quad \mu'_4 = E[X^4] = \int_{-\infty}^{\infty} x^4 f(x) dx$

We expect a normal distribution to have skew and kurtosis (Fisher's definition) that are very close to zero. In addition to the skew and kurtosis, hypothesis testing via the Shapiro-Wilk[3] test can provide evidence for or against normality. The Shapiro-Wilk statistic, $W$, is computed by:

$$W = \frac{\left(\sum_{i=1}^{n} a_i x_{(i)}\right)^2}{\sum_{i=1}^{n}(x_i - \bar{x})^2}$$

where $a_i$ are constants generated from the mean, variances, and covariances of the order statistics of a sample $n$ that come from a normal distribution, $x_{(i)}$ is the $i^{th}$ order statistic, and $\bar{x}$ is the sample mean. From this statistic, we computed the p-value and performed the hypothesis test, where the null hypothesis is that the distribution of the sample is normal.

### Sampling with Python

For efficient computation, we used the SciPy[4] statistics package and the NumPy[5] random package in Python. SciPy's statistics package includes a large variety of statistical distributions, tests, etc. that are easy to use for statistical modeling and analysis. Likewise, NumPy offers pseudo-random number generation and sampling from a wide variety of distributions. SciPy was used to compute the skew, kurtosis, and the Shapiro-Wilk test statistic and p-value, while NumPy was used to generate random samples from the normal and Laplace distributions.

## Results

We found strong evidence against the assumption that the daily returns follow a normal distribution both numerically and graphically. Calculation of the skew, kurtosis, and Shapiro-Wilk test statistic and associated p-value revealed that the daily returns of the major stock indices deviate from the expected properties of a normal distribution (Table 1). Using a standard normal distribution as a benchmark, the differences in the skew, kurtosis, test statistic (W), and p-value are obvious.

Table 1 - **Measures of Skew, Kurtosis, and Normality in Historical Return Data.** The standard normal distribution was used as a benchmark comparison for the distribution properties of the daily stock returns. From top to bottom, sample sizes for the computations are N = 5000, 1879, 1879, and 1879. All data for stock returns were taken from Yahoo Finance and extend back to January 2012.

| Sample | Skew | Kurtosis | W | p – value |
|---|---|---|---|---|
| Standard Normal Distribution | 0.0273 | -0.1446 | 0.9995 | 0.2256 |
| S&P 500 Index | -0.3431 | 3.2465 | 0.9557 | 1.9151E-23 |
| Dow Jones Industrial Average | -0.3480 | 3.4969 | 0.9564 | 2.8225E-23 |
| NASDAQ Composite | -0.3527 | 2.7256 | 0.9626 | 1.2728E-21 |

Table 1 demonstrates the non-normality of the distributions of daily stock returns, but as we saw in Figure 1, the distributions are rather symmetric. Fitting the data to normal and Laplace distributions, we compared the empirical cumulative distribution function (ECDF) of the data with the cumulative distribution function (CDF) of the fitted models (Figure 2). We see that the normal approximation does not include the outliers, and it also does not fit the rest of the data as well as the Laplace approximation. The Laplace both captures the outliers and has a better overall fit to the data.

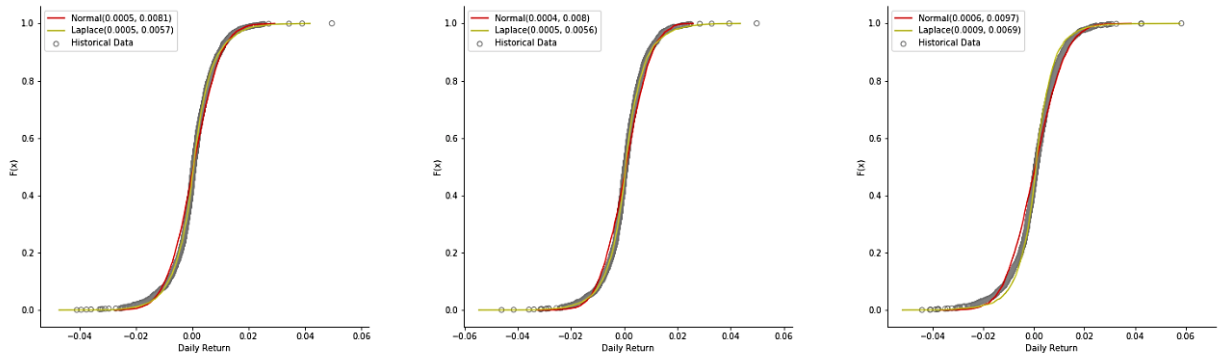

Figure 2 – **(Left to Right) Empirical Cumulative Distribution Functions of the S&P 500 Index, Dow Jones Industrial Average, and the NASDAQ Composite with Normal and Laplace Approximations.** Normal and Laplace models were fitted to the data of each of the three stock indices, and their CDFs, $F(x)$, were plotted against the ECDFs of the raw data.

## Conclusion

Our investigation of the daily stock returns of major stock indices shows that using a Normal approximation to model returns leaves out a large amount of valuable information about the data, particularly the chance of observing a very positive or very negative return in a single day. Numerical and graphical representations of this fact force us to look beyond the normal model. We offer the Laplace distribution as a good alternative to model daily returns because it is symmetric and has fat-tails. This distribution is more fitting to stock market returns, and it allows professionals working in the industry to get a more realistic expectation of daily returns.

However, as we have said before, no model is perfect. We suggest that future work explores extreme-value distributions or fat-tail distributions that provide a similar framework as the Normal distribution – symmetric – and the Laplace distribution – symmetric and fat-tailed. Furthermore, the probability density function of the Laplace distribution is continuous but not differentiable at its center, which allows for further study in mathematically well-behaved functions to model the behavior of stock returns. That being said, the Laplace model is a first-step into more accurate models of daily stock returns, and it or some other extreme-value distribution provides a much more robust description of stock market data.